\documentstyle[aps,prl]{revtex}
\begin{document}
\bibliographystyle{unsrt}

 \newcommand{\um}[1]{\"{#1}}
 \newcommand{\uck}[1]{\o}
 \renewcommand{\Im}{{\protect\rm Im}}
 \renewcommand{\Re}{{\protect\rm Re}}
 \newcommand{\ket}[1]{\mbox{$|#1\protect\rangle$}}
 \newcommand{\bra}[1]{\mbox{$\protect\langle#1|$}}
 \newcommand{\proj}[1]{\mbox{$\ket{#1}\bra{#1}$}}
 \newcommand{\expect}[1]{\mbox{$\protect\langle #1 \protect\rangle$}}
 \newcommand{\inner}[2]{\mbox{$\protect\langle #1 | #2
\protect\rangle$}}

\title{On energy transfer by detection of a tunneling atom}

\author{A.M. Steinberg}

\address{Department of Physics, University of Toronto \\
Toronto, ONT M5S 1A7 CANADA}

\date{\today}
\maketitle

\begin{abstract}
We are in the process of building an experiment to study the 
tunneling of laser-cooled Rubidium atoms through an optical barrier.
A particularly thorny set of questions arises when one considers
the possibility of observing a tunneling particle while it is
in the ``forbidden'' region.  In earlier work, we have discussed how
one might probe a tunneling atom ``weakly,'' so as to prevent collapse.
Here we make some observations about the implications of a more
traditional quantum measurement.  Considerations of energy 
conservation suggest that attempts to observe tunneling atoms will
enhance inelastic scattering, but not in a way which can be directly
observed.  It is possible that attempts to make such measurements
may lead to experimentally realizable 
``observationally assisted barrier penetration.''
\end{abstract}

 \vskip1pc

\section{Introduction}

Tunneling is one of the most striking predictions of quantum 
mechanics, and continues to provoke some of its most heated 
controversies.  Despite the appearance of tunneling phenomena in 
numerous physical and technological areas (and in first-year physics 
courses), certain aspects of the effect remain poorly understood.  
This is seen most clearly in the debate over how long a particle 
takes to traverse a tunnel barrier, and in particular, whether or not 
it can do so faster than light 
\cite{Chiao=1997ProgOpt,Steinberg=1998AnnPhys}.

The confusion over these issues can be traced to certain common 
elements of quantum ``paradoxes.''  For one, definite trajectories 
cannot be assigned to particles in general, and in this sense it is 
not even clear how to rigorously phrase a question about how much time 
a transmitted particle spent in a forbidden region-- in fact, it may 
not even be necessary that a particle ``traverse'' a region in order 
to be found on the far side.

Of course, at one level quantum mechanics
is merely a wave theory, and quite thoroughly understood.  
In many physical situations, more controversial, interpretational, 
issues (related to ``collapse'' or other alternatives) may easily be 
skirted without loss of predictive power.  In tunneling, however, it 
is quite natural to look for a description of transmitted particles, 
as distinct from reflected ones (or from the ensemble as a whole) 
\cite{Buttiker=1982}.  But such a description is impossible without an 
attempt to model the detection itself, because without the detection 
event, transmitted and reflected packets necessarily coexist.  Detection
naturally raises other interesting questions.  What is the nature of a
detection process which occurs {\it inside} a forbidden region
(cf. \cite{Carniglia=1972JOSA,Carniglia=1971PRD})?  According to the collapse postulate, if a particle
is found to be in the barrier region, it is subsequently described by a
new wave function, confined to that region.  Any such wave function
has $E>V_0$, and suddenly, the problem is no longer one of tunneling.

Our plans to observe tunneling of laser-cooled Rubidium atoms, and 
to perform ``weak measurements''\cite{Aharonov=1990,Aharonov=1988} in order to study the 
behaviour of a transmitted subensemble, have been presented at length 
elsewhere\cite{Steinberg=1998Superlatt,Steinberg=1998AnnPhys,Myrskog=1999}.  Here we repeat 
only the essential elements, to provide context for the present discussion.

\section{Tunneling in atom optics}

Laser-cooled atoms offer a unique tool for studying quantum phenomena 
such as tunneling through spatial barriers.  They can routinely be 
cooled into the quantum regime, where their de Broglie wavelengths 
are on the order of microns, and their time evolution takes place in 
the millisecond regime.  They can be directly imaged, and if they are 
made to impinge on a laser-induced tunnel barrier, transmitted and 
reflected clouds should be spatially resolvable.  With various 
internal degrees of freedom (hyperfine structure as well as Zeeman 
sublevels), they offer a great deal of flexibility for studying the 
various interaction times and nonlocality-related issues.  In 
addition, extensions to dissipative interactions and questions related 
to irreversible measurements and the quantum-classical boundary are 
easy to envision.\cite{Steinberg=1998Superlatt}

In our work, we prepare  a
sample of laser-cooled Rubidium atoms in a MOT, and cool them in
optical molasses to
approximately 6 $\mu K$.  As explained below, further cooling techniques are 
under investigation for
achieving yet lower temperatures \cite{Myrskog=1999}.

We plan to use a
tightly focussed beam of intense light detuned far to the blue of the 
D2 line to create a dipole-force potential for the 
atoms\cite{Rolston=1992,Miller=1993FORT,Davidson=1995}.  In this
intense beam, the atom becomes polarised, and the polarisation lags
the field by $90^{\circ}$ when the light frequency exceeds that of
the atomic resonance.  This polarisation out of phase with the local
electric field constitutes an effective repulsive potential, 
proportional
to the intensity of the perturbing light beam.  It can also be thought
of in terms of the new (position-dependent) energy levels of the atom
{\it dressed} by the intense laser field.
Using a 500 
mW 
laser at 770 nm, we will be able to make repulsive potentials with 
maxima on the order of the Doppler temperature of the Rubidium vapour.
Acousto-optical modulation of the beam will let us shape these 
potentials with nearly total freedom, such that we can have the atoms 
impinge on a thin plane of repulsive light, whose width would be on 
the order of the cold atoms' de Broglie wavelength.  This is because 
the beam may be focussed down to a spot several microns across 
(somewhat larger than the wavelength of atoms in a MOT, but of the 
order of that of atoms just below the recoil temperature, and hence
accessible by a combination of cooling and selection techniques).  
This focus may be rapidly displaced  by 
using acousto-optic modulators and motorized mirrors.  As the atomic motion is in the 
mm/sec range, the atoms respond only to the time-averaged intensity, 
which can be arranged to have a nearly arbitrary profile.

As a second stage of cooling, we follow the MOT and optical molasses
with an improved variant of a proposal termed ``delta-kick 
cooling''\cite{Ammann=1997}.  In our version, the millimetre-sized 
cloud is allowed to expand for about ten milliseconds, to several times 
its initial size.  This allows individual atoms' positions $x_{i}$
 to become strongly correlated 
with atomic velocity, $x_{i} \approx v_{i}t_{\rm free}$.  Magnetic 
field coils are then used in either a quadrupole or a harmonic 
configuration to provide a position-dependent restoring force for
a short period of time.  By 
proper choice of this impulse, one can greatly reduce the rms velocity
of the atoms.  So far, we have achieved a 
one-dimensional temperature of about 700 nK, corresponding 
to a de Broglie wavelength of about half a micron.  We are currently 
working on improving this temperature by producing stronger, more 
harmonic potentials, and simultaneously providing an antigravity potential 
 in order to increase the interaction time.

However, the tunneling probability through a 5-micron focus will 
still be negligible at these temperatures.  Furthermore, the 
exponential dependence of the tunneling rate on barrier height will 
be difficult to distinguish from the exponential tail of a thermal 
distribution at high energies.  We will therefore follow the 
delta-kick with a velocity-selection phase\cite{Myrskog=1999}.  
By using the same beam which is to form a tunnel barrier, but 
increasing the width to many microns, we will be able to ``sweep'' 
the lowest-energy atoms from the center of the magnetic trap off to the 
side, leaving the hotter atoms behind. Our simulations suggest
that we will be able to
to transfer about 7\% of the atoms into the one-dimensional ground
state of this auxiliary trap.  This new, 
smaller sample will have a thermal de Broglie wavelength of 
approximately $3.5 \mu$m, leading to a significant tunneling 
probability through a 10-micron barrier.  We expect rates on 
the order of 1\% per secular period, causing the auxiliary trap to 
decay via tunneling on a timescale of the order of 100 ms. 

\section{Measuring tunneling atoms}

A weak 
measurement is one which does not significantly disturb the particle 
being studied (nor, consequently, does it provide much information on 
any single occasion).  Why not perform a strong measurement?  Simply 
because if one can tell with certainty that a particle is in a given
region, one has also determined that the particle has enough energy to
be in that region; one is no longer studying tunneling.  The 
measurement has too strongly disturbed the unitary evolution of the
wave function.

At the 6th Symposium on Laser Spectroscopy in Taejon, I made the above 
glib assertion as I had frequently done in the past, and went 
on to discuss weak measurements.  
Afterwards, however, Bill Phillips raised the question of {\it where} 
exactly the energy comes from to turn a forbidden region into an 
allowed one.  The imaging of an atom involves a small transfer of 
momentum, and typically the only energy exchange is the more-or-less 
negligible recoil shift.  But in this scenario, an atom observed under
an arbitrarily high tunnel barrier must-- merely by being observed--
 acquire enough energy 
to ride on top of the barrier.  Why should the effect of a
weak probe beam (in fact, the interaction with a single resonant 
photon) scale with the completely unrelated height of a potential 
barrier?

The situtation envisioned is shown in schematic form in Fig. 1.  The wave
function of the atom decays exponentially into the barrier region over a
characteristic length $1/\kappa$.  If this length is greater than the resolution
of the imaging lens, then it is possible for the appearance of a spot of
focussed fluorescence on an appropriate point on the screen to indicate
that an atom is in the barrier region.  This atom, having scattered perhaps
only a single photon, must according to quantum mechanics have acquired
an energy of at least $V_0$ to exist confined to the barrier region.  This energy
depends not on the wavelength or intensity of the imaging light, but only on
the height of the barrier created by the dipole-force beam, which may greatly
exceed the recoil energy associated with the momentum transfer involved
in elastic scattering of a photon.  An interesting point is that it is unnecessary to
actually focus and detect the photon in question.  The very possibility that some
future observer {\it could} use the scattered light to determine that an atom was 
in the forbidden region is sufficient to decohere spatially separated 
portions of the atomic wavefunction, causing some fraction of the 
atoms to ``collapse'' (if you will) into the barrier region.

At first, one might think that the energy comes from the interaction between the
atom and the dipole-force beam.  A little thought suffices, however, to demonstrate
that this cannot be the solution.  Even in the absence of a potential, an imaging
beam may localize a previously unlocalized particle, increasing its momentum 
uncertainty and hence its energy.  The energy must come from the imaging beam.
Why, then, does the quantity of energy transferred depend on the barrier height?

A partial answer comes from carefully considering the energy levels of the atom.
Inside the barrier region, the presence of the dipole beam couples the atomic
eigenstates, creating an AC Stark shift
(which is the effective repulsive potential).  An atom
which makes a transition between a state primarily outside the barrier (of energy
$E_g + P^2/2m$) to a state localized in the barrier is simultaneously making a
transition to a new, higher-energy electronic state 
($E_{g}+V_{0}+P^{\prime\, 2}/2m$).  Energy-conservation will be enforced
by the time-integral in perturbation theory, causing the amplitudes for this process
to interfere destructively unless the scattered photon energy plus the final energy of
the atom equals the initial photon energy plus the initial energy of the atom.  In
other words, the presence of the dipole beam makes possible inelastic (Raman) transitions
between different atomic states.  When an elastic scattering event occurs, the atom is
left in the original state, and cannot be localized to the barrier.  Only when an inelastic
collision occurs can the atom be transferred to the state dressed by the dipole field,
and localized in the formerly forbidden region.

Can one then determine that an atom is in the barrier {\it without} imaging, by merely 
measuring the energy of the scattered photon?  Unfortunately, no.  Recall that this argument
hinges on an imaging resolution 
\begin{equation}
\delta l < 1/\kappa \;, 
\end{equation}
where 
\begin{equation}
\hbar^2\kappa^2 = 2m(V_0-E) \; .
\end{equation}
A particle localized to within $\delta l$ has a momentum uncertainty 
\begin{equation}
\Delta P \geq \hbar/2\delta l\; .
\end{equation}
This means that it will only remain within the resolution volume for a time
\begin{equation}
t \leq \frac{\delta l}{\Delta P/m} = 2m\delta l^2/\hbar\; .
\end{equation}
  This in turn implies
\begin{equation}
t < 2m/\hbar\kappa^2\; .
\end{equation}
 Unless the imaging light is time-resolved to better
than this limit, it is impossible to maintain the spatial resolution necessary to conclude
with certainty that the particle is in the barrier.  (Strictly 
speaking, it would suffice to image to better than the barrier width.
However, a particle in the barrier is most likely to be within the 
first
exponential decay length $1/\kappa$.  While with lower resolution, 
one might still conclude that the particle was deeper in the barrier, 
the likelihood will be exponentially suppressed.  Thus on some 
occasions, the photon energy will be shifted by an amount greater 
than its rms spectrum, but the low amplitude of this frequency 
component will be matched by the low probability of finding the atom 
so deep in the barrier, and the present arguments may easily be 
generalized.)  
This implies that the energy uncertainty
of the scattered photon must be 
\begin{equation}
\Delta E \stackrel{>}{_\sim} \hbar/t > \hbar^2\kappa^2/2m \; .
\end{equation}
But this is $V_0-E$, just the energy required to excite the tunneling atom above the barrier.
So the only way to image an atom in the forbidden region is to use light with sufficient energy
uncertainty that it can boost the atom above the barrier {\it without} a significant change in
its own spectrum. 

\section{Detection without observation, or observation without 
detection?}

Just as these issues were beginning to make themselves clear to us, Terry Rudolph 
suggested an even more confounding extension.  His idea is outlined in 
Figure 2.
Suppose one decides to determine the location of the tunneling atom in a more indirect
manner.  Specifically, suppose a nearly ideal imaging system is devised (relying, for
example, on $\pi$-pulses of probe light), but that a beam stop is imaged onto the barrier
region.  In this way, any atom in the classically allowed region will be imaged, but an
atom which finds itself in the forbidden region will be out of the reach of probe
light, and no photon will be scattered.  When no scattered photon is observed,
we can conclude with near certainty that the atom is in the forbidden region, and
has therefore made a transition to a higher-energy dressed state.  But now where
did the energy come from?  After all, it appears that the ``detected'' atom became
localized without ever undergoing an interaction.

This picture is ill-founded, however.  The atom cannot be considered in isolation; it is in fact
the entire system, composed of atom, dipole-force beams, and imaging photons,
which undergoes a transition and must conserve energy.  Under the influence of a
probe pulse, the atom's quantum state becomes entangled with the state of the
imaging light.  There is some amplitude for a photon to travel along its original path,
unscattered, but this amplitude is correlated with an atomic state localized to the
barrier region.  For the time-integral of this amplitude to lead to a real probability
for detecting an unscattered photon, the detected photon will necessarily lose enough
energy to boost the atom above the barrier, just as in the case previously discussed.

Once more, the situation becomes less startling when we observe that 
(1) there is indeed a mechanism for energy-exchange between the ({\it 
unscattered}) imaging beam and the atom; and (2) this energy exchange 
never exceeds the intrinsic uncertainty in the initial photon energy.
The interaction between imaging light and an atom comprises not only 
the possibility of scattering, but also the real part of the atomic 
polarizability, which is to say the index of refraction experienced 
by the light due to the presence of the atom.  For a near-resonant 
photon with a probability $\eta$ of being scattered by an atom, the 
extra optical path introduced by the presence of the atom, $\int 
[n(z)-1]dz$, is of the order of an optical wavelength times $\eta$, 
corresponding to an optical phase shift approximately equal to $\eta$.
If an atom is found to have appeared in the dark region enclosing the 
barrier, this implies that it left the region of interaction with the 
probe light, causing the light to experience a time-varying index of 
refraction.  If the atom's departure from the illuminated region is 
known to have occurred within a time $t$, then the phase of the light 
was modulated by an amount $\eta$ in a time smaller than $t$, 
producing a frequency shift of the order $\eta/t$.  Each photon's 
energy can in this way be altered by the ``disappearance'' of the 
atom, by the quantity $\hbar\eta/t$.  Since on the order of $1/\eta$
photons are necessary to detect atoms with near-unit probability in 
such a scenario, this phase-modulation effect is automatically 
sufficient to produce an energy exchange of up to $\hbar/t$ between 
the moving atom and the probe beam, {\it even when no photons are 
scattered}.

As in the original discussion of {\it bright} imaging of the barrier
region, we know that $\hbar/t \stackrel{>}{_\sim} 
\hbar^{2}\kappa^{2}/2m$, and this energy exchange is enough to propel
the particle above the barrier.  Furthermore, the same argument 
concerning
the duration of a probe pulse remains intact.  If the pulse lasts long 
enough that even a particle localized to the barrier would have time 
to escape while the light was on, then one will never completely avoid 
fluorescence, and thus never be able to conclude with certainty that 
the particle is in the barrier region.  One might instead envision a 
CW probe but time-gated photodetection; in this case, the argument is 
similar, but it is the {\it detected} photon whose energy can no 
longer be determined precisely enough to be certain that energy 
exchange has taken place on any individual occasion.  Nevertheless, by 
studying an ensemble of particles, one should be able to build up 
enough statistics to confirm the shift in the mean photon frequency.

\section{Conclusion}
We see that tunneling is just one more prototypical example of the 
way in which observation may disturb a quantum system.  It is 
instructive to consider the mechanisms which allow the necessary 
energy transfer to take place, along with the requisite uncertainties
behind which this transfer hides.  Ultracold atoms in Bose 
condensates, and at temperatures achievable through related laser-cooling techniques 
as well, have long enough de Broglie wavelengths that tunneling effects
should soon be observed in a regime where these questions become
more than purely academic.  Particularly intriguing is the 
possibility of {\it modifying} the barrier-traversal rate by the 
application of a probe beam which {\it could} in principle be used
to image an atom in the forbidden region.  Even if no attempt is made 
to actually perform the imaging, the simple possibility that one 
could do so should be enough to turn the quantum amplitude for an
atom to be within about $1/\kappa$ of the edge of the barrier into
an actual probability, in the sense of a real fraction of atoms
localized into that region of the barrier.  These atoms have enough
energy to traverse the barrier classically in either direction, and
may therefore be observed on the far side.

\section{Acknowledgments}
This discussion would remain purely academic if not for the hard work
of Stefan Myrskog, Jalani Fox, Phillip Hadley, and Ana Jofre
on our laser-cooling experiment.
I would also like to acknowledge Jung Bog Kim and his students Han Seb Moon 
and Hyun Ah Kim for their collaboration on this project.  
I want to thank Jung Bog Kim and the organizers of the Symposium on Laser 
Spectroscopy for their invitation and for their hospitality during the 
meeting, which proved quite stimulating.  Finally, I would like to 
thank Bill Phillips for the question which prompted this short paper, 
and for fascinating discussions concerning it; and Terry Rudolph for 
following it up with an even harder question just as I thought I was 
beginning to understand something.

\newpage

{\bf Figure Captions}
1. In this setup, a tunneling atom is illuminated by a plane wave, and
the scattered fluorescence may be imaged on a screen to determine 
whether or not the particle was in the barrier region.

2. Here, a beam block is imaged onto the barrier region, in such a way
that an image may be observed on a screen {\it unless} the particle is 
in the process of tunneling.

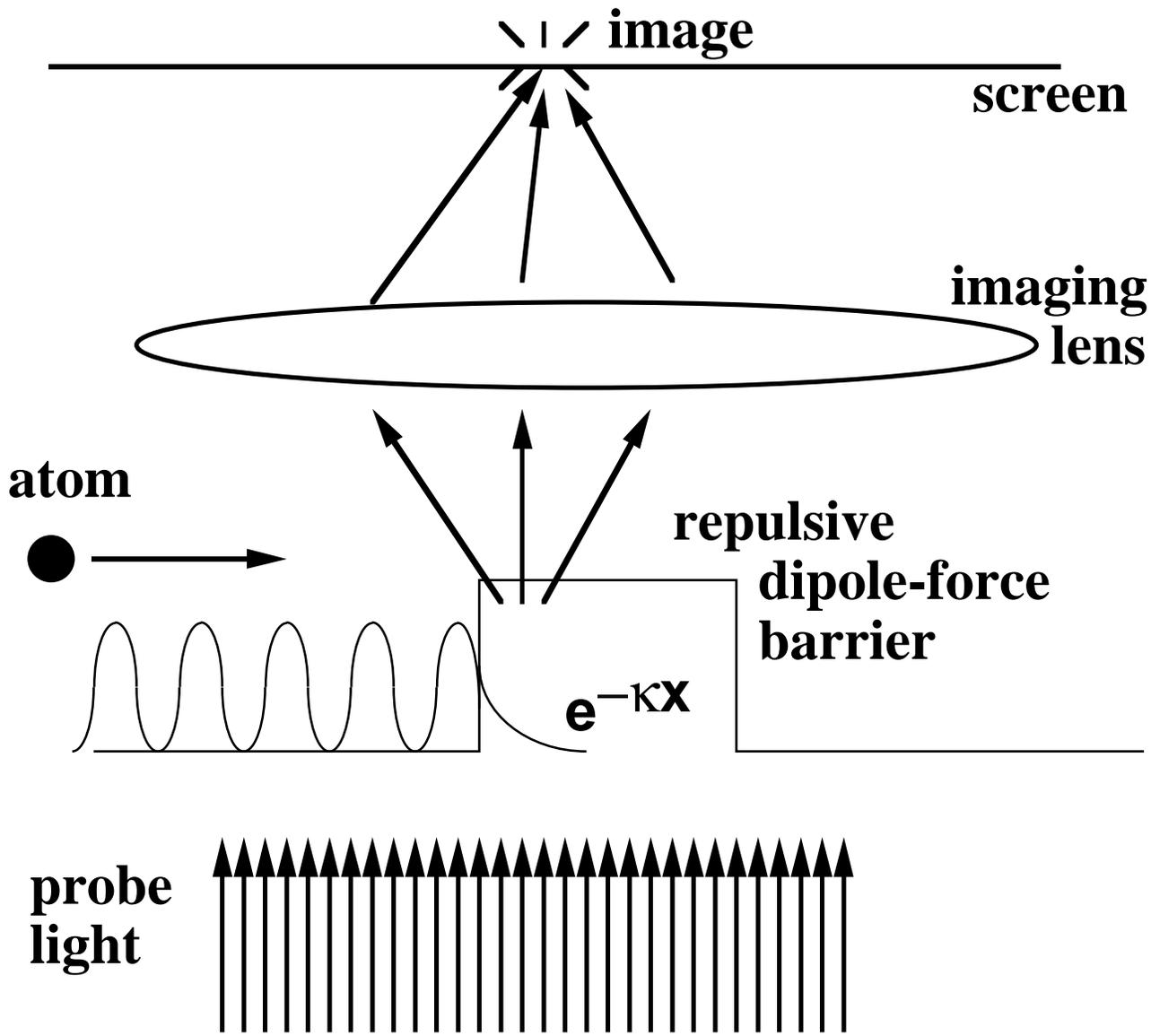
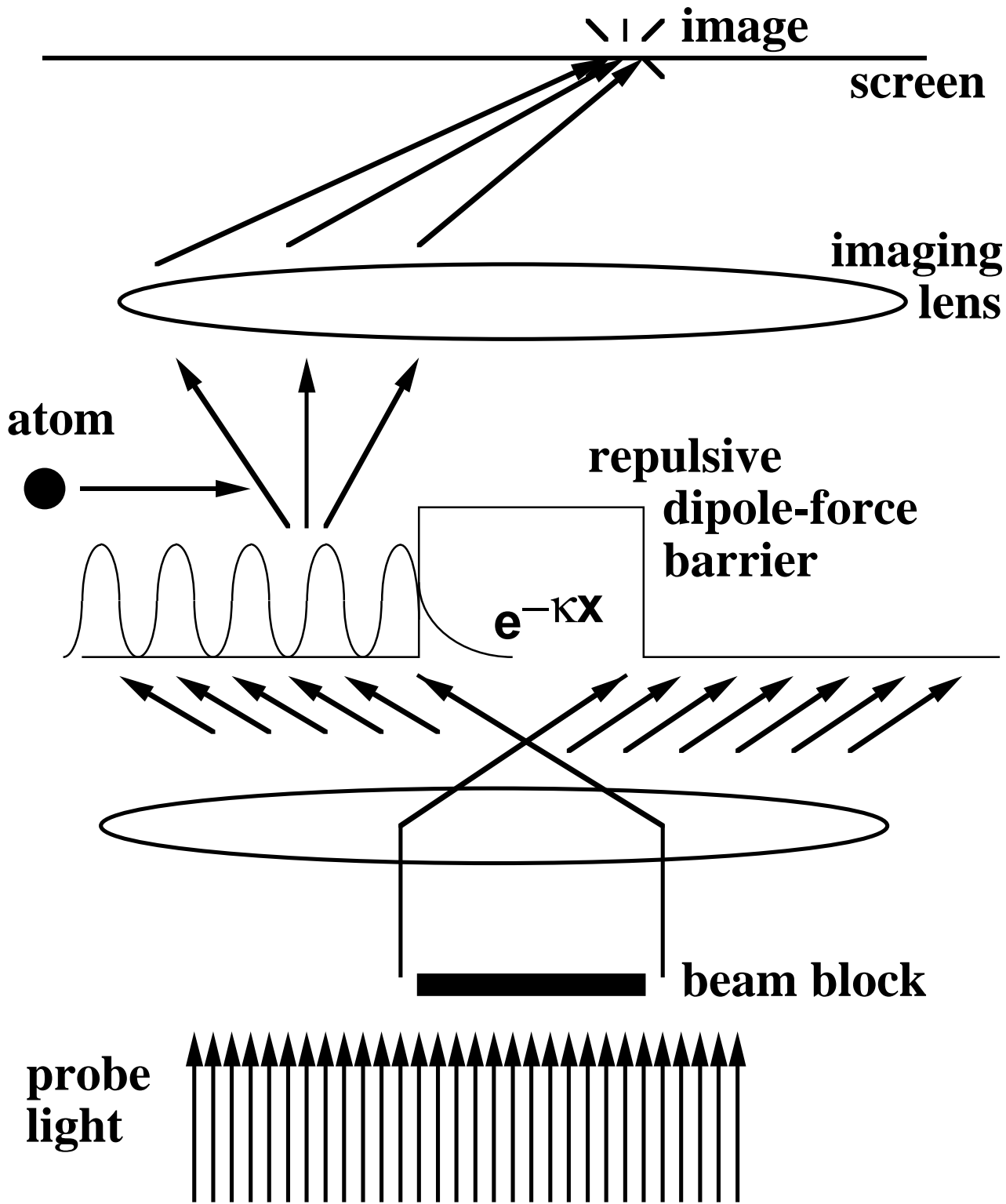
\end{document}